\documentclass[sigconf,authorversion,nonacm,screen]{acmart}

\AtBeginDocument{%
  \providecommand\BibTeX{{%
    \normalfont B\kern-0.5em{\scshape i\kern-0.25em b}\kern-0.8em\TeX}}}

\usepackage{soul}
\usepackage{svg}
\usepackage{balance}

\setcopyright{rightsretained}
\acmPrice{15.00}
\acmDOI{10.1145/3573382.3616052}
\acmYear{2023}
\copyrightyear{2023}
\acmSubmissionID{playcomp23pop-p1006-p}
\acmISBN{979-8-4007-0029-3/23/10}
\acmConference[CHI PLAY '23]{Companion Proceedings of the Annual Symposium on Computer-Human Interaction in Play}{October 10--13, 2023}{Stratford, ON, Canada}
\acmBooktitle{Companion Proceedings of the Annual Symposium on Computer-Human Interaction in Play (CHI PLAY '23), October 10--13, 2023, Stratford, ON, Canada}
\received{2023-06-22}
\received[accepted]{2023-08-03}

\begin{document}

\title[Link, User-Centred Designer]{Link, User-Centred Designer:\\ Game Characters as Transcendent Models}

\author{Katie Seaborn}
\email{seaborn.k.aa@m.titech.ac.jp}
\orcid{0000-0002-7812-9096}
\affiliation{%
  \institution{Tokyo Institute of Technology}
  \city{Tokyo}
  \country{Japan}
}

\renewcommand{\shortauthors}{Seaborn}

\begin{abstract}
  Games allow us to construct and explore identities and offer us role models, good and bad. Game characters are a reflection of us--players and creators alike--or could be. But do games also encode identities, values, and orientations that transcend diegetic categories and player self-insertion? I explore the notion of game characters as conduits of transcendent models through the case study of Link from the Legend of Zelda series. I propose that designers embed tacit, nondiegetic patterns of praxis and complex value models, such as user-centred design, when crafting the embodiment of characters in gameplay, even unawares.
\end{abstract}



\begin{teaserfigure}
  \includegraphics[width=\textwidth]{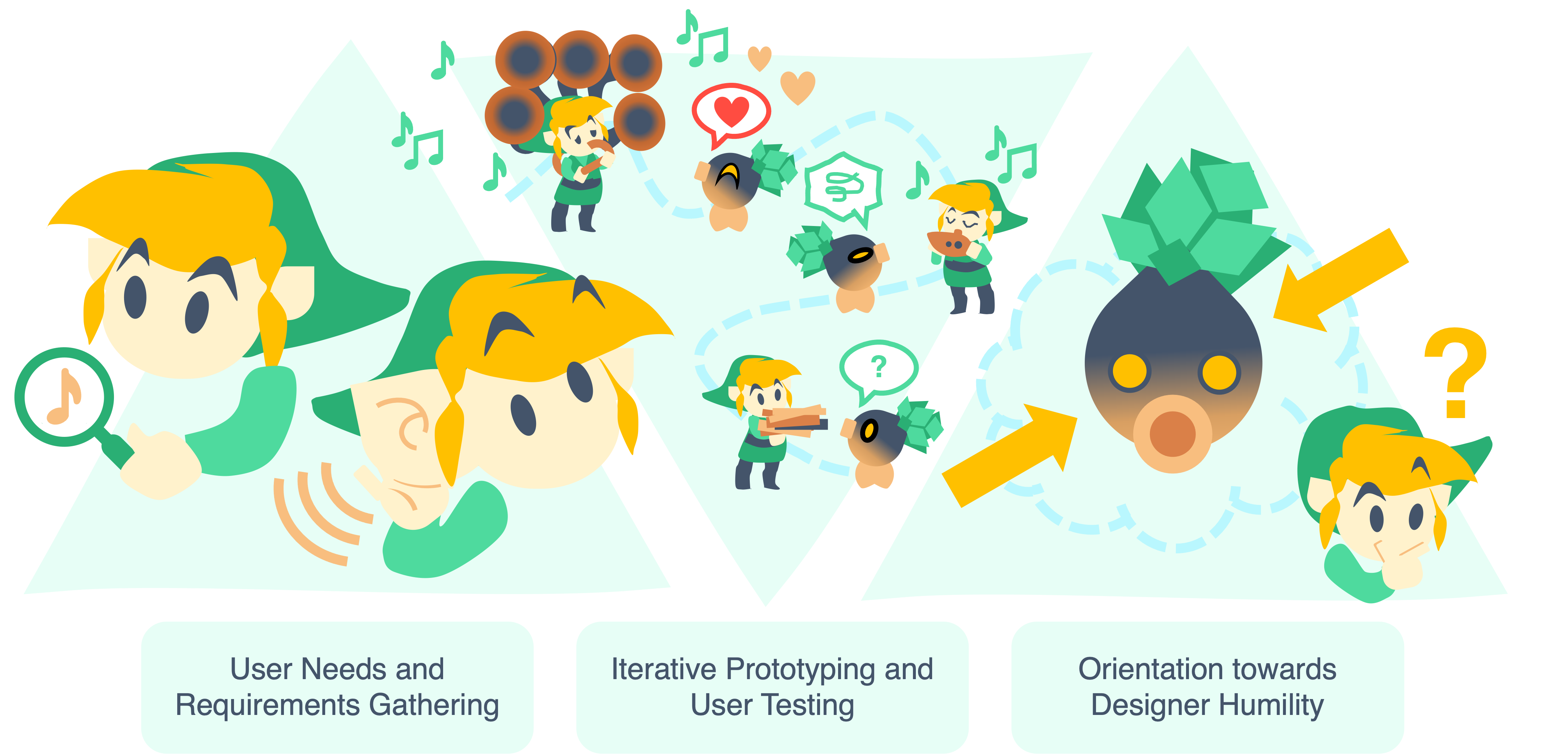}
  \caption{Link embodies the transcendent model of user-centered designer.}
  \Description{Three panels depicting example scenarios in which Link embodies fundamental aspects of user-centred design praxis: needs and requirements gathering; prototyping and user testing; and designer humility.}
  \label{fig:teaser}
\end{teaserfigure}

\received{20 February 2007}
\received[revised]{12 March 2009}
\received[accepted]{5 June 2009}

\maketitle

\section{Introduction}
When you think of Link, hero of \emph{The Legend of Zelda} series, what comes to mind? Perhaps his bravery as holder of that piece of the Triforce. Or the Hylian sword and shield, liberally employed on game packages and merchandise. Some may simply think "that little green guy in a toque" (or beanie, for Americans). A legion of players, young and old, casual and hardcore, have played as the speechless (but not voiceless\footnote{Link showcases the difference between speech and voice. Link never speaks out loud, yet frequently lets out vocal bursts \cite{brooks_deep_2023}: grunts, exclamations, cries. Link verbalizes replies to conversational prompts in text, but are these words Link's, the player's, or something else?}) Link over several offerings.\footnote{I exclude the cartoon series and Japanese comic, which lean on rather non-canonical representations of Link.} Many have experienced the wonder and joy of venturing through the worlds of Hyrule \emph{as} Link. The "as" here is critical: we control Link \emph{and} take on his identity within the game space and as long as the magic circle holds. Link is a part of us, even briefly, an experience we can return to time and again, with or without an ocarina.

The relationship between the player and the played is well-trodden ground. Konijin and Bivjank \cite{kojinin_doors_2009} describe how we construct identities with and through game characters. Turkle \cite{turkle_ghosts_1995} considered how simulations and interactive media, notably virtual worlds and games, beget \emph{identity fragmentation}, or, more positively, \emph{identity fluidity}. In games, this is constrained by the gameplay and narrative trappings, unlike in open worlds and sandboxes. Single-player games---most of the \emph{The Legend of Zelda} series--are further limited. The social element is lost; identity construction is bound by what is offered within the game space. This may seem a harrowed blow, as identity construction is a social process \cite{tajfel_social_1978, tajfel_social_2004, hogg_social_2016}. Still, people can experience one-sided parasocial relationships \cite{horton_mass_1956}: a psychological merging of player and character \cite{little1999manifesto,lewis2008they}, where the player conforms to the identity of the character, i.e., \emph{the Proteus effect} \cite{yee2007proteus}, or projects themself onto the character, forming a \emph{blended} \cite{harrell2013phantasmal} or \emph{projected} identity \cite{gee2003video}. There can be divergence, where the player invests in the character and feels a part of their social world \cite{hartmann_parasocial_2008,banks2015object}. Game characters may also act as role models \cite{merton_unanticipated_1936}: representations of ideal selves that we aspire to be. Link has been iconic in this regard. Even as a masculine character partaking in violent acts, Link also frequently engages in feminine-coded acts of kindness and helpfulness, and is heralded for his bravery, perhaps a gender-neutral value. Coupled with his ambiguous appearance, this liminal characterization has provided room for gender-diverse folks to see themselves in Link and experience gender euphoria \cite{codega_link_2023}. These orientations point to a boundary, albeit a porous one, between "being" the character and "being with" the character.

Far less explored is how creators---character design is usually a team effort, despite our propensity to single out a celebrity designer---embed themselves within the characters that they create. By this, I do not suggest creator self-insertion, nor awareness on their part, even while I frame this as an act of non-diegesis. What I refer to is the creators "being in" the character. I argue that this is (i) a feature of the \emph{creation process}, perhaps unintentional and unavoidable, and (ii) reflected in the \emph{embodiment} of the character within the gameplay. By embodiment, I mean how the character \emph{through} their body interacts with objects and others in the environment \cite{miller_embodiment_2016}. The use of the game medium is also key. This is not merely bias or "slant," nor is it necessarily negative \cite{friedman_bias_1996}. Like Flanagan and Nissenbaum \cite{flanagan_values_2014}, I recognize that people embed values of all kinds in the games that they create, knowingly or otherwise. I wish to add another branch to the tree: that creators embed \emph{complex value models} that are \emph{intersectional} and \emph{pluralistic}, such that they cannot be reducible to a single value. I call this a \emph{transcendent model}, as such patterns transcend simple categories \emph{and} the diegeis of the game world.

\section{Case Study of Link: From Hero to Human-Centred Designer}
Link is an ideal candidate: a positive force embraced by generations of players across a suite of games created by Nintendo, a company known for its player-centred creation process\footnote{\url{https://shmuplations.com/mario64/}} and playtesting regimen. I argue that Link embodies a transcendent model pertinent to the CHI PLAY community: \emph{user-centred designer}\footnote{I use the terms user-centred and human-centred interchangeably and for stylistic purposes.} \cite{international_standards_organization_ergonomics_2016}. I justify the concept by outlining\footnote{I use zero subheadings to bring home the meta, transcendent nature of this idea.} \st{four swords} three key examples of how Link embodies this transcendent model, drawn from typical and time-worn player experiences across the \emph{The Legend of Zelda} games.

\subsubsection{User Needs and Requirements Gathering}
Link is inclined to offer help to non-playable characters (NPCs).
These side quests are often based on seeking out the "user" needs of the NPC. Link "interviews" other NPCs and "observes" the "context of use" or "user probes" for clues that could relate to these "user needs," such as snooping in Kafei’s diary at his mother's blessing (\emph{Majora's Mask}). Link can be given an initial "list of user requirements," anything from recipes and shopping lists to ill-defined "desires" and "problems" that require Link to flesh out by more deeply "centring" the NPCs as individuals who may or may not differ in various ways from Link himself. In a classic example, Link realizes that he must learn the favoured song of horse Epona's original owner to calm the beast's nerves (\emph{Ocarina of Time}).

\subsubsection{Prototyping and User Testing}
Link is a prototyper and iterative designer. This is best represented in the mechanics of \emph{Tears of the Kingdom} and \emph{Breath of the Wild}. Vehicles, mecha, contraptions, cages, cooking recipes ... Link can craft, test, and re-craft "prototypes" for himself and others. Earlier games provided a simpler process of Link scrounging around and offering solutions to problems posed by NPCs, "iterating" until the NPC was satisfied. Never self-serving, this orientation towards others represents a link between worlds, the NPC or "user" and the solution or "technology."\footnote{By "technology," I do not necessarily mean the thing created, but rather the \emph{application} of knowledge in a repeatable solution to a problem \cite{skolnikoff_elusive_1993}.}

\subsubsection{Designer Humility}
Link is humble and inclusive. He does not foist solutions and worldviews on others. Even though he is a Hylian (most of the time), he does not lean on that worldview or restrict his efforts to those of his race. In \emph{Majora's Mask}, masks are empathy suits \cite{lavalliere_walking_2017}, allowing Link to take on the embodiment--not just the appearance, but also the interactive capacities and social situatedness--of other Hyrulian races, from the Goron to the Deku, and even specific individuals or "personas," such as Mikau, a Zora rock band guitarist. Still, he literally wears his positionality on his sleeve, or head, in his iconic hat. And, as always, he is all but silent, listening rather than speaking.

\subsection{Counterpoint and Foil: Ganon}
Ganon is Link's nemesis, but also presents a foil for the transcendent model of user-centred designer. Ganon is all about domination and achieving self-serving desires, the antithesis to user-centredness, unless we consider him the sole "user." In \emph{Ocarina of Time}, Ganon persuades the Gerudos, by trick or force, into doing things his way over a participatory approach: deceptive design \cite{brignull2023decept} at its finest. Ganon is compelling as a character, but do we designers also find a little bit of ourselves in him? He may model immanence that deserves dismantling through \emph{brave} reflexivity \cite{rode2011reflex}.

\section{Conclusion}
Link is a venerated character and a case study of how transcendent models representing complex nondiegetic values can be found in game creations. Whether creators have done this on purpose or faithfully embody these models themselves is up to debate. Discovering what other transcendent models exist is another adventure.

\begin{acks}
Thank you to Giulia Barbareschi and Brandon Sichling for pre-reviewing this paper. I also thank the chairs and the external peer reviewers for specially handling and reviewing this paper.
\end{acks}

\bibliographystyle{ACM-Reference-Format}
\balance
\bibliography{main.bbl}

\end{document}